Title: United States FDA drug approvals are persistent and polycyclic: Insights into economic cycles, innovation dynamics, and national policy

Author: Iraj Daizadeh, PhD, Takeda Pharmaceuticals, 40 Landsdowne St. Cambridge, MA, 02139, iraj.daizadeh@takeda.com

Abstract: It is challenging to elucidate the effects of changes in external influences (such as economic or policy) on the rate of US drug approvals. Here, a novel approach – termed the Chronological Hurst Exponent (CHE) – is proposed, which hypothesizes that changes in the long-range memory latent within the dynamics of time series data may be temporally associated with changes in such influences. Using the monthly number FDA's Center for Drug Evaluation and Research (CDER) approvals from 1939 to 2019 as the data source, it is demonstrated that the CHE has a distinct S-shaped structure demarcated by an 8-year (1939-1947) Stagnation Period, a 27-year (1947-1974) Emergent (time-varying Period, and a 45-year (1974-2019) Saturation Period. Further, dominant periodicities (resolved via wavelet analyses) are identified during the most recent 45-year CHE Saturation Period at 17, 8 and 4 years; thus, US drug approvals have been following a Juglar/Kuznet mid-term cycle with Kitchin-like bursts. As discussed, this work suggests that (1) changes in extrinsic factors (e.g., of economic and/or policy origin ) during the Emergent Period may have led to persistent growth in US drug approvals enjoyed since 1974, (2) the CHE may be a valued method to explore influences on time series data, and (3) innovation-related economic cycles exist (as viewed via the proxy metric of US drug approvals).

Keywords: FDA approvals, drug development, medical product, medicines, economic cycle, Schumpeter

**Introduction**

Drug discovery and development (DDD) requires investment to maneuver a single putative medicine from discovery science to market approval for a given condition or disease. The investments cover the costs associated with acquiring both the hardware (e.g., laboratory materials and space) and software [explicit (e.g., patents) and tacit (e.g., know-how) knowledge] as well as executing the various DDD activities [1]. Ultimately, should an investigational candidate survive the attrition process and obtain marketing authorization (also known as marketing approval) by a health authority, a sponsor then enjoys economic rents secured from supplying the approval medicine. On the demand side, the patient receives a trusted medicine associated with a market innovation based on a new chemical and biologic entity, a cost advantage (generic), or a more efficient delivery of drug product [2].

Since the early 20$^{th}$ century to the present, in terms of drug development, the social, economic, and political environments have evolved dramatically. For example, the growth in the amount of governmental investment in research and development (R&D) [3], the number of R&D firms [4, 5], the volume of intellectual property (e.g., patents, trademarks, as well as peer-reviewed publications) [5, 6], the number of R&D policy initiatives (see Table 1 and discussion below), and the rise of the R&D cluster [7] have seemingly grown synchronistically and exponentially. As a case in point, in the US and across industries, Daizadeh [8,9] showed a statistical significant intercorrelation in the time course of R&D investment, the number of patent and trademark applications, peer-reviewed and media publications, and stock price of major indices in the US.

Importantly, the 20$^{th}$ century also gave rise to the modern regulated DDD industry including the invention of an objective, independent, and external agency (collectively termed the health authority (HA)). The HA performs a vital function by attesting to a medicine's quality, safety, and efficacy profile and to formally authorize a drug for marketing purposes in a given jurisdiction. Since its original conception, there have been increasing refinement in its scope proportional to changes in the social environment through amendments in policy. For example, focusing on the US Food and Drug Administration (FDA), there has been an evolution in the number and variety of policy initiatives focused on providing oversight to the DDD process due exclusively to important social concerns regarding safety and efficacy of certain drugs circulating in inter-jurisdictional commerce [38]. The FDA policy environment has evolved considerably from placing under regulation on specific drugs (e.g., insulin and penicillin) and describing the basic tenets of the safety sciences in its infancy to building a robust infrastructure commencing in the 1960s with the Kefauver-Harris amendments to regular updates in the policy landscape starting in 1977, with the introduction of the Bioresearch Monitoring Program, pushing the frontiers of regulatory science into the $21^{th}$ century (see Table 1; [38]).

Concomitantly, economic factors have also greatly influenced the landscape of DDD. Unlike the US (see 21 CFR 310 *et seq.*; 21 CFR 601 *et seq.*), in many jurisdictions (e.g., the European Union), HAs consider cost and/or reimbursement when assessing the merits of granting a marketing application. The ability of sponsors to obtain the economic rents from supplying quality, safe and efficacious HA-authorized drugs is a key driver that has sustained the DDD process. Among other factors, expected revenues from marketing an innovative HA-approved drug product would be proportional to monopolizing power of the intellectual property [1] as well as the amount of labor required to move the drug from concept to delivery, thereby requiring a broad assortment of various investments in terms of tangible and intangible assets. While beyond the scope of this work, cost estimates to secure marketing authorization vary based on the types of challenges experienced in various phases (e.g. target / modality in discovery; the number, length and type of clinical trials in development) [41], with significant savings expected with expediting development [42]. Thus, drug approvals may be thought of an economic outcome within a given jurisdiction, and should behave as such. One such test would be to investigate the presence of cycles in the number of approvals similar to that found in other forms of economic output.

Economic cycles, a wavelength between crests of development maxima over stagnation minima, are an active area of inquiry, not without controversy [10]. Juglar defined this periodicity over three phases: prosperity, crisis, and subsequent liquidation, and suggested an "approximate length of the cycle with crisis/liquidation taking 1-2 years, followed by a 6-7 year phase of prosperity [11; pp. 7]," with drivers to prosperity to crisis transition due to exuberance and thus over-speculation (ibid). Kitchin derived 'minor' and 'major' inventory cycles with wavelengths of 3.5 years (40 month) and "aggregates usually of two, and less seldom of three, minor cycles," respectively [12; pp. 10]. Subsequent to the introduction of these short and intermediate cycles, Kondratieff introduced the concept of the long-wave 50-60 year cycles [13]. Concomitantly, Kuznetz extrapolated 15-25 year cycles derived from data from "fluctuations in rates of population growth and immigrating but, also with investment delays in building, construction, transport infrastructure, etc… [14; pp. 2]." These authors extrapolated the information from a broad assortment of macro-economic data from US and Europe including climate, monetary, fiscal, consumption, among others.

Memory characteristics (also termed persistency) in the dynamics of typical econometrics captured over time are intimately connected with cycles and thus also to the underlying processes [15]. Technically, however, these same characteristics such as long-range memory processes are challenging to analyze and interpret due to (in part) self-similarity and typical non-stationary properties (as they confound spurious from true signals) [16]. The Hurst constant and wavelet analyses are statistical time series tools that may be calculated in such as a way as to avoid these challenges [17]. While there are other ways to define a Hurst constant, a measurement of memory, it is classically defined as $H \sim \ln(R/S)_t / \ln(t)$, where R and S is the rescaled range and standard deviation, respectively, and t is a time window. An H=0.5, an H<0.5, and an H>0.5 indicates a random walk, an anti-persistent, and a persistent (trend reinforcing) time series, respectively [18]. Wavelet analyses is a well-established group of time series methods that leverages the expansion and contraction of wave functions to resolve time series properties [19].

In this work, and the to the author's knowledge, this is the first investigation of the existence and evolution of persistency, and the existence of approval cycles (akin to economic cycles) within US drug approvals, which is treated as a macro-economic variable and a proxy metric for FDA policy. This work is exploratory and empirical in nature. As presented in the Materials and Methods section below, the data source is a time series of monthly values of US drug Approvals from Jan. 1939 through Dec. 2019 from the Centers of Drug Evaluation and Research (CDER) branch of the Food and Drug Administration (FDA), which "regulates over-the-counter and prescription drugs, including biological therapeutics and generic drugs[1]." While this is not the only institution that regulates the DDD process within the FDA, it is one that provides a publicly, reliable and valuable source of longitudinal metrics regarding the DDD process from the dawn of the review process (1939) to the present time. The methods are standard with the exception of the Chronological Hurst Exponent to explore the persistency latent in the time series. All datasets and R Project code are provided in the Electronic Supplementary Materials section for the sake of transparency and replicability as well as to encourage future researchers in investigate a potentially very interesting and informative aspect of drug development. This work then discusses the key results of both the descriptive and inferential statistics followed by a discussion on how the statistical work positively supports the hypotheses mentioned above (viz., persistency and economic cycles are latent within US drug approvals), and the ramifications of this work including potential linkages to sociological, economic, and policy features experienced over the nearly 100 years of data.

**Materials and Methodologies**

The following summarizes the data sources and the statistical approaches used. This work is applied by nature and thus differing the mathematical formulae and technical discussion to original sources, as cited. All data and the R Project code for the statistical analysis are provided in the Electronic

---

[1] https://www.fda.gov/about-fda/fda-organization/center-drug-evaluation-and-research-cder

Supplementary Materials section supporting this article for transparency and reproducibility, as well as for purposes of future work.

*Data Sources and Data Preparation*

The data was obtained from the FDA repository accessed at https://www.accessdata.fda.gov/scripts/cder/daf/ on July 16 and July 17, 2020. The data was culled from a monthly report and described as follows:

> "All Approvals and Tentative Approvals by Month.
>
> Reports include only BLAs/NDAs/ANDAs[2] or supplements to those applications approved by the Center for Drug Evaluation and Research (CDER) and tentative NDA/ANDA approvals in CDER. The reports do not include applications or supplements approved by the Center for Biologics Evaluation and Research (CBER).
>
> Approvals of New Drug Applications (NDAs), Biologics License Applications (BLAs), and Abbreviated New Drug Applications (ANDAs), and supplements to those applications; and tentative approvals of ANDAs and NDAs."

Upon entry into the data-repository via the website, the number of approvals from Jan. 1939 to Dec. 2019 was then determined by month. The values were placed in Excel and then exported as a comma delimited comma-separated values (CSV) file for input into the data analysis routine.

The total dataset comprised 181,157 total approvals from Jan. 1939 until Dec 2019 (for a total of 972 monthly observations). The author notes that submission history for each approval during this roughly 100 year time-period was not found on the US FDA website.

---

[2] BLAs/NDAs/ANDAs: Biologics License Applications, New Drug Applications, Abbreviated New Drug Applications

*Statistical Analysis*

As mentioned above, as this is an applied paper, reference is made to the various theoretical formulae in the respective supportive citations. Many of the distribution-inquiring statistical tests selected are considered 'standard' in the sense that they are typically used in the context described and are readily available and interpretable. All methods presented below followed standard implementation; default parameters were used (as appropriate) throughout the analyses. While the R code [20] is presented in the Electronic Supplemental Materials section of this article, the steps to perform the analysis were as follows:

I. Load US Approvals as a time-series and perform descriptive statistics (including autocorrelation functions) [21; R package: 'moments'].

In this step, the data is read as a time series into the R program, and descriptive statistics including moments and serial and partial correlation functions calculated.

II. Assess attributes of the time series, including:

- Normality [22; R package: 'nortest'] using the Anderson-Darling and Cramer-von Mises normality tests
- Stationarity [23; R package: 'aTSA'] using the Kwiatkowski-Phillips-Schmidt-Shin (KPSS) Unit Root Test for both the original and single difference
- Long-memory [24; R package: 'LongMemoryTS'] using the Qu and local Whittle score tests
- Seasonality [25; R package: 'seastests'] using the WO, QS, Friedman and Welch tests
- Nonlinearity [26; R package: 'nonlinearTseries'] using the Teraesvirta's and neural network tests, and Keenan, McLeod-Li, Tsay, and likelihood ratio tests.

III. Determine the Chronological Hurst Exponent (that is, evaluate if the Hurst exponent over time evolves):

For a given time-series, the Hurst constant [27; R package: 'tsfeatures'] is a statistical indicator of the memory in a time-series process (or processes). In this calculation, the time-varying nature of the H constant was investigated using time windows from the first datapoint (Jan. 1939) to the end of the window length, with 1-month increments. The algorithm to calculate the Chronological Hurst Exponent is as follows:

```
hurstApprovals=0; end<-length(time)
for (I in 1:end) { hurstApprovals[i]  <- hurst (time[1:(1+i*1)]) }
hurstApprovals<-ts(hurstApprovals,start=c(1939,1),end=c(2019,12),frequency=12)
```

IV. Determine the periodicities within the time series:

Wavelet analyses used to investigate the structure of the periodicities within the time series given its dynamics (particularly its non-stationarity; see step II). Two wavelet methods were utilized: one with a smoothing (Loess) approach [28; R package: 'WaveletComp'] and one [29-31; R Package: 'dplR'] without. The average period versus the average power for each method was then calculated to elucidate the main periodicities. The dominant frequency was then re-checked with spectral analysis [32-33; R Package: 'forecast'].

**Results**

*Descriptive statistics: Elementary properties of the chromodynamics of US drug approvals*

The time series of US drug approvals follows an interesting flow given the dramatic rise starting in the 1970s to 2000 then after a drastic fall with a subsequent re-rise (Figure 1).

< Insert Figure 1 here. **Figure 1: Time evolution of total US CDER Approvals** >

The US drug approvals time series distribution is non-normal, platykurtic and positively skewed, with an average of 186 approvals (191 standard deviation) (Table 2 and 3).  Importantly, the time series is non-stationary, non-seasonal, and non-linear, with intrinsic persistent memory (Table 2 and figure 2), which is removed with single differencing (that is, the time series has an order of integration (number of differences to attain stationarity) of 1, I(1)). I(1) processes are rather well-represented across a spectrum of different disciplines and a broad assortment of the economic variables including US drug approvals [34].

< Insert Table 2 here: **Table 2: Descriptive statistics of US approvals (rounded to tenths; units in months)** >

< Insert table 3 here: **Table 3: Summary of tests investigating normality, stationarity, seasonality, long-memory, and non-linearity**>

< Insert Figure 2 here: **Figure 2: Serial and partial correlation functions: lag is presented in months** >

*Chronological Hurst Exponent: Existence of economic cycles and latent persistency*

Using the Chronological Hurst Exponent approach to investigate the long-term memory processes of the time-series shows, interestingly, a unique trichotomized structure (Figure 3). Three periods are clearly shown: Period 1: prior to June 1947, a period of stagnation with H~0.5; Period 2: June 1947 to May 1974, a period of time-varying nature (also herein called emergent), where the H constant fluctuates rises under a degree of fluctuation; and, Period 3: May 1974 to Dec 2019, a period of saturation in which the H~1.

< Insert Figure 3 here: **Figure 3: The Chronological Hurst Exponent based on US Drug Approvals (Figure 1) from 1939 to 2019**>

Concordantly, the wavelet periodogram during Period 3 demonstrates that the time series contains periodicities. Several relatively long-, medium-, and short-range periodicities are observed during this period: 16-18 years (with a maximum (black ridge) occurring at 17 years), ~4-8 years, and on the monthly, yearly, or biyearly periodicities presenting intermittently, respectively (Figure 4). The predominate periodicity is identified to be 17, 8 and 4 years from spectral analysis (Figure 5).

< Insert Figure 4 here. **Figure 4: Wavelet periodogram of US approvals: black lines are the wavelet power ridges and white contour lines to border the area of wavelet power significance of 99%** >

< Insert Figure 5 here. **Figure 5: Wavelet period versus power with 95% significant levels in red** >

**Discussion and Conclusion**

Using time series analysis, this work finds two conceptually novel aspects of US drug approvals: the existence and evolution of persistency, and the existence of approval cycles (akin to economic cycles).

*Persistency*

Formally, persistency may be defined as the "rate at which its autocorrelation function decays to zero," or "the extent to which events today have an effect on the whole future history of a stochastic process [40]." Translating to the context of our concern, it generally means that the value of US drug approvals at a given month is closely related to its value at the prior month. The Chronological Hurst Exponent proposed herein is a simple algorithm that reiteratively calculates the Hurst exponent (a measure of persistency) over an incrementally increased time period. With each iteration, an additional data point (here the next monthly observation of US approvals) is taken into account until the exponent of the full data set is calculated. The Chronological Hurst Exponent proposed in this work elucidated a S-shaped structure reflecting a trichotomized picture of the time evolution of persistency latent within US drug approvals:

- Period 1: An 8-year (1939-1947) stagnation period in which the Hurst exponent remained at or around 0.5. An Hurst exponent at these values suggest no persistency whatsoever.

- Period 2: A 27-year (1947-1974) time-varying (emergent) period in which the Hurst exponent gradually evolved from 0.5 to 0.9. This range in the Hurst exponent suggests a growing persistency within the time series data.

- Period 3: A 45-year (1974-2019) saturation period in which the Hurst exponent remained at or around 1. A saturated Hurst exponent implies that the time series has become (for lack of a better term) inelastic; that is, any further changes in the degree and/or number of exogenous variables do not affect the persistency course of the time series (as it is already maximized).

*Cyclicity*

Interpreting US drug approvals as an economic variable – a singular outcome of several complex macro- (national), meso- (cluster), and micro (firm)-inputs such as national policy and R&D spend (government, firm), potential of future rents (individual buyer, payor), science and technology innovation (tacit (e.g., staff dexterity) and explicit (e.g., patents) knowledge), and resource availability (e.g., chemicals, vials) – the existence of business cycles were investigated. Several tiered periodicities (17 years, 4-8 years, and intermittent monthly/yearly) were identified within Periods 2 and 3 of the CHE. Thus, one of the key findings of this work is that approval cycles, similar to economic cycles, exist. These approval cycles seem to be the result of explanatory variables that are working in a cumulative manner.

*Persistence and Cyclicity Interpreted*

During Period 2 (27-years (1947-1974)), it is observed that 1947 was the first year in which there were one or more approvals during much of the year and had the largest number on an annual basis since the start of the collection cycle in Jan 1939. After 1947, a general rise in the number of approvals per month and per annum is observed. It is also a period of commensurate changes to the policy and social landscape pertaining to DDD, as well as continued investment into R&D. These changes were seemingly due to end of World War II (1939-1945), the beginning of the so-called 'Golden Age of Capitalism,' and the associated economic progress [35] with a relatively small number of economic disasters (see Figure 3 in [36]). Since the 1938 Food, Drug and Cosmetic act, no significant advances in policy occurred until the 1962 Drug Amendments (see Table 1), while there were significant milestone activities in terms of congressional review (the Kefauver Hearings dealt with pricing and market control [37]. One could therefore speculate that it an overall increased economic activity (and not specific FDA policy changes, per se) that may have driven the changes in the persistency measurement.

The appearance of Period 3 (45-years (1974-2019)) suggests a uniform pressure onto the time series. Two general reasons present themselves to foment such a sustained persistent alteration in the fabric of US drug approvals: some sort of substantive and everlasting change (1) to accounting practices regarding US drug approvals (that is, how the source data was initially contrived and/or collected); or (2) in the scientific, social, economic, and/or legislative landscape. The former is unlikely to cause a persistent shift. To illustrate, FDA data sources state a change in department ownership in and around that time, as well as issues regarding changes from fiscal to calendar year practices.[3] It is unlikely that either of these reasons would have changed the time series in such a permanent manner. The latter reason, while likely, however, is ill-defined, but does allow for hypothesis generation.

One hypothesis that could be tested is that of a significant change in the FDA policy landscape (see Table 1) may have caused the formation of Period 3. From an FDA perspective, the 1960s and 1970s were a transformative vicennial [38]. In 1962, the Kefauver-Harris (KH) amendments to the original Food, Drug and Cosmetics Act (FD&C) of 1938 introduced (inter alia) broad requirements on drug efficacy (including key concepts of 'substantial evidence' and 'adequate and well-controlled studies'), drug quality (via good manufacturing practices), ethical guidelines (patient informed consent), and physician-researcher supervision of the clinical trials. Subsequently, a review of prior-to-1962 approved drugs were retrospectively investigated based on the evidentiary standard of the KH amendments, which led to revocation of "over 1000 ineffective drugs and drug combinations from the marketplace (page 13 of *ibid*)." The concepts such as those introduced in the KH amendments (partly listed above) have been refined and reinforced through ongoing congressional action, directly contributing to the identified persistency affect and cyclicity. Ongoing policy actions, such as Prescription Drug User Fee Act (PDUFA) and its subsequent 5-year amendments commencing in 1992, or the introduction of new technologies may have directly contributed to innovation-based periodicities, leading to significant increases in the promulgation of guidelines that may have furthered drug approvals [34, 39].

---

[3] Data record information from https://www.fda.gov/about-fda/histories-product-regulation/summary-nda-approvals-receipts-1938-present (extracted on July 30, 2020).

Thinking outside of the drug development process and continuing considering the periodogram (Figure 5) and thinking of the original time series (Figure 2), the complex periodicity profile may have been motivated by socio-economic factors. Substantive economic pulses that may have affected the overall approval flow may include: Black Monday Market Crash (October 19, 1987), the Dot-Com bubble burst (Q3, 2002), and the subprime mortgage crisis (September 17, 2008), among others. Visually, the Dot-Com bubble burst seemed to coincide with a downsizing of amplitude. However, it is difficult to ascertain if the other triggers may have affected the time series.

Interestingly, if one considered the US drug approvals strictly as an economic variable, and assuming the theory of Schumpeter's economic cycles, the identified periodicities seem to coincide with certain macro-economic periodicities, with exception as no canonical long-term (> 40 years) periodicities were identified in this analysis (see Table 4). The periodicities began at different times with different durations (Figure 4). The dominant periodicity of 17, 8 and 4 years has reoccurred during the longest (45 years), medium (20 years), and short-term (intermittent) durations, respectively (Figure 5). Thus, it seems that US drug approvals follow a Juglar/Kuznets mid-term cycle with Kitchin bursts. Only time will tell if a longer-term cycle (Kondratieff) emerges, irrespective of any downside pressures (such as multi-decade bear cycles). A key difference between the identified approval cycles as compared with economic cycles may be the degree of importance of the regulatory context. While a potentially coarse interpretation, without the legal requirement for market approval there would not have been a US drug approvals time series, whereas for variables such as gross domestic product typically used to consider economic cycles this is not the case (as the legal regimes do not define (as much as support) the existence of these more traditional economic variables).

< Insert table 4 here. **Table 4: Mapping of broad canonical economic cycles with that of periodicities associated with US Approvals** >

*Further Thoughts in Light of Limitations of Current Study*

There are extensions and limitations to any statistical analyses, especially when dealing with social-economic variables. Examples of future investigation may include:

Hypothesis:

- One could argue that the number of US drug applications may have been a more insightful variable, as applications may be either withdrawn (by the Sponsor) or rejected (by the FDA). Unfortunately, the author could not find this dataset.

- The number of initial US drug applications or approvals for new molecular and/or biologic entities may provide additional insight into the economics of the innovative process. In this article, the total number of US drug approvals including generics and line extensions (e.g., new indications or dosage forms) were considered, as reflected "market innovation." That is, a sponsor would not have considered seeking an approval without a market driver of some sort.

Data:

- Data integrity and completeness: This study relies on a single source dataset from the FDA. While the author feels comfortable with the data source, there is uncertainty in how the data is

- collected, maintained, and presented given the duration of data collection and limited-to-no ability to cross-reference.

- Data transformation: The data was transformed from irregular to a regular time-structure. That is, FDA drug approvals occurred as a function of day; these data were then aggregated into monthly values to facilitate the statistical analyses. Thus, some information may have been lost in terms of structure, as there are limited statistical routines able to manage such data.

In the author's opinion, these data are an important artifact of R&D expenditures related to the DDD industry and therefore have interesting utility. Future investigations may consider these data and analyses to support research questions such as those related to forecasting and long-memory effects of non-stationary and non-linear data. It will be interesting to revisit these analyses on a yearly basis given the recent COVID-19 crises and resultant economic challenges, with a hope that the US drug approvals remain persistent with respect to these significant triggers.

Study Conclusions:

In conclusion, this work introduces the Chronological Hurst Exponent, an algorithm which examines the time evolution of long-term memory intrinsic to time series data. Using this algorithm, US drug approvals are examined. The CHE of US drug approvals is found to follow a distinctive S-shaped (trichotomized) curve, with three periodicities that seem to be correlative with the evolving US drug development policy landscape, as well as macro-variable changes that may be relevant to drug development. Further, using wavelet analysis, cyclicity in the frequency of US drug approvals was observed in the most recent period identified in the CHE analysis. These periodicities adds evidence to the concept of mid-term economic cycles, assuming US drug approvals data are viewed a proxy metric of innovative capacity. The empirical findings and statistical approaches outlined in this report promise an exciting new frontier of further research into the various forces driving drug development.

**Acknowledgements**

The author extends gratitude N.D., S.L.D., and N.L.D. for their support of the manuscript.

**Disclosures**

The author is an employee of Takeda Pharmaceuticals; however, this work was completed independently of his employment. The views expressed in this article may not represent those of Takeda Pharmaceuticals. As an Associate Editor for Therapeutic Innovation and Regulatory Science, the author was not involved in the review or decision process for this article. See Electronic Supplementary Materials for all data and methods to replicate (or extend) the results presented herein.

Persistency and cyclicity in US drug approvals                                    Author: Daizadeh, I.(39) Daizadeh, I. (2020) Since the Mid-2010s FDA Drug and Biologic Guidelines have been Growing at a Faster Clip than Prior Years: Is it Time to Analyze Their Effectiveness? Ther Innov Regul Sci. https://doi.org/10.1007/s43441-020-00233-0

(40) Caporale, Guglielmo Maria and Pittis, Nikitas, Persistence in Macroeconomic Time Series: Is it a Model Invariant Property?. Revista de Economia del Rosario, Vol. 4, No. 2, pp. 117-142, 2001, Available at SSRN: https://ssrn.com/abstract=928506

(41) Chorniy, A., Bailey, J., Civan, A. and Maloney, M. (2021), Regulatory review time and pharmaceutical research and development. Health Economics, 30: 113-128. https://doi.org/10.1002/hec.4180

(42) DiMasi, J.A., Wilkinson, M. The Financial Benefits of Faster Development Times: Integrated Formulation Development, Real-Time Manufacturing, and Clinical Testing. *Ther Innov Regul Sci* **54,** 1453–1460 (2020). https://doi.org/10.1007/s43441-020-00172-w
Page 21 of 45

**Table 1: Brief Milestones in FDA Drug Regulation [Daizadeh, 2020][4].**

| Year | US Drug Regulation |
|---|---|
| 1938 | Act and Requirements for Premarket Drug Safety and New Labeling |
| 1941 | The Insulin Amendment |
| 1945 | The Penicillin Amendment |
| 1951 | Durham-Humphrey Amendment |
| 1962 | Kefauver-Harris Drug Amendments |
| 1977 | Introduction of the Bioresearch Monitoring Program |
| 1981 | Revision of the regulations for human subject protections |
| 1982 | Tamper-resistant Packaging Regulations issued |
| 1983 | Orphan Drug Act |
| 1984 | Drug Price Competition and Patent Term Restoration Act (Hatch–Waxman Act) |
| 1987 | Investigational drug regulations |
| 1988 | FDA Act of 1988 and Prescription Drug Marketing Act |
| 1989 | Guidelines on significant use in elderly people |
| 1991 | Accelerated review of drugs for life-threatening diseases; Common Rule adopted across agencies |
| 1992 | Generic Drug Enforcement Act; co-establishes International Conference on Harmonization (ICH); Prescription Drug User Fee Act (PDUFA I) |
| 1993 | MedWatch launched; revising women of childbearing potential in early phase drug studies policies and assessments of genders-specific medication responses |
| 1994 | Uruguay Round Agreements Act |
| 1995 | Cigarettes as 'drug delivery devices' |
| 1997 | FDA Modernization Act (FDAMA); reauthorization of PDUFA II |
| 1998 | Adverse Event Reporting System (AERS); Demographic Rule; Pediatric Rule |
| 1999 | ClinicalTrials.gov; guidances for electronic submissions; drug facts; Prescription Drug Broadcasting Advertising Final Guidance; Managing the Risks from Medical Product use: Risk Management Framework published |
| 2000 | Data Quality Act |
| 2002 | Best Pharmaceuticals for Children Act; Public Health Security and Bioterrorism Preparedness |

---

[4] https://www.fda.gov/about-fda/virtual-exhibits-fda-history/brief-history-center-drug-evaluation-and-research

| | |
|---|---|
| | and Response Act of 2002; Current good manufacturing practice (cGMP) initiative; PDUFA III; outcomes of pregnancies registries guidance |
| 2003 | Medicare Prescription Drug Improvement and Modernization Act; Pediatric Research Equity Act |
| 2004 | Project BioShield Act of 2004; Anabolic Steroid Control Act of 2004; "Innovation or Stagnation?—Challenge and Opportunity on the Critical Path to New Medical Products" published; bar code introduced |
| 2005 | Drug Safety Board announced; risk management performance goal guidances |
| 2006 | Requirements on Content and Format of Labeling for Human Prescription Drug and Biological Products final rule |
| 2007 | PDUFA IV; FDA Amendments Act (FDAAA) |
| 2008 | Sentinel Initiative |
| 2009 | FDA Transparency Initiative |
| 2010 | FDA Transparency Results Accountability Credibility Knowledge Sharing (TRACK) |
| 2012 | PDUFA V; Launch of FDA Adverse Event Reporting System (FAERS); Food and Drug Administration Safety and Innovation Act (FDASIA); Generic Drug User Fee Amendment |
| 2013 | Drug Quality and Security Act; Mobile Medical Applications; Global Unique Device Identification Database (GUDID) |
| 2016 | 21st Century Cures Act |
| 2017 | Current Good Manufacturing Practice (cGMP) Requirements for Combination Products; FDA Reauthorization Act (FDARA; PDUFA VI) |

**Table 2: Descriptive statistics of US approvals (rounded to tenths; units in months)**

| Minimum | 1st Quartile | Median | Mean | Standard Deviation | 3rd Quartile | Maximum | Kurtosis | Skew |
|---|---|---|---|---|---|---|---|---|
| 0 | 5.0 | 164 | 186.4 | 190.9 | 392.2 | 858 | 2.6 | 0.7 |

**Table 3: Summary of tests investigating normality, stationarity, seasonality, long-memory, and non-linearity**

| Test Category | Test Name | Test statistic | Outcome against null hypothesis |
|---|---|---|---|
| Normality | Anderson-Darling test | p-value < 2.2e^16 | Normal distribution rejected |
|  | Cramer-von Mises test | p-value < 7.37e-10 |  |
| Stationarity | KPSS unit root test* | 0.01 (for no drift/no trend; for drift/no trend; for drift/trend) | Stationarity rejected |
| Long memory | Qu test* | 1.033545 versus 1.517 (alpha=0.01;eps=0.02) | Long memory accepted |
|  | Multivariate local Whittle Score* | 1.668473 versus 1.517 (alpha=0.01) |  |
| Seasonality | Webel-Ollech test | p-value 0.05 | "The WO-test does not identify seasonality" |
|  | QS test, Friedman, Welch tests |  | False – seasonality rejected |
| Linearity | Teraesvirta's neural network test | p-value=0 | Linearity in "mean" rejected |
|  | White neural network test | p-value=0 | Linearity in "mean" rejected |
|  | Keenan's one-degree test | p-value=3.889e^-5 | The time series follows some AR process rejected |
|  | McLeod-Li test | p-value=0 | The time series follows some ARIMA process rejected |
|  | Tsay's test | p-value=6.45e^-14 | Time-series follows some AR process rejected |
|  | Likelihood ratio test for threshold non-linearity | p-value=0.0004552571 | Time-series follows some TAR process rejected |
| * Some tests require stationary data.  As such, as the number of differences required for a stationary series from the original time-series was 1, the difference was used in the specific test demarcated. ||||

**Table 4: Mapping of broad canonical economic cycles with that of periodicities associated with US Approvals**

| Theory | Periodicity | US Approvals |
|---|---|---|
| Kitchin Short-Term Cycle Cycle | 3.5 years | Months to biannual |
| Juglar Mid-Term Cycle | 7-11 years | 4-8 years |
| Kuznets Medium-Term Cycle | 15-25 years | 17 years |
| Kondratieff Long-Term Cycle | 40-60 years | |

**Figure 1: The number of monthly US CDER Approvals as a function of year from 1939 to 2019**

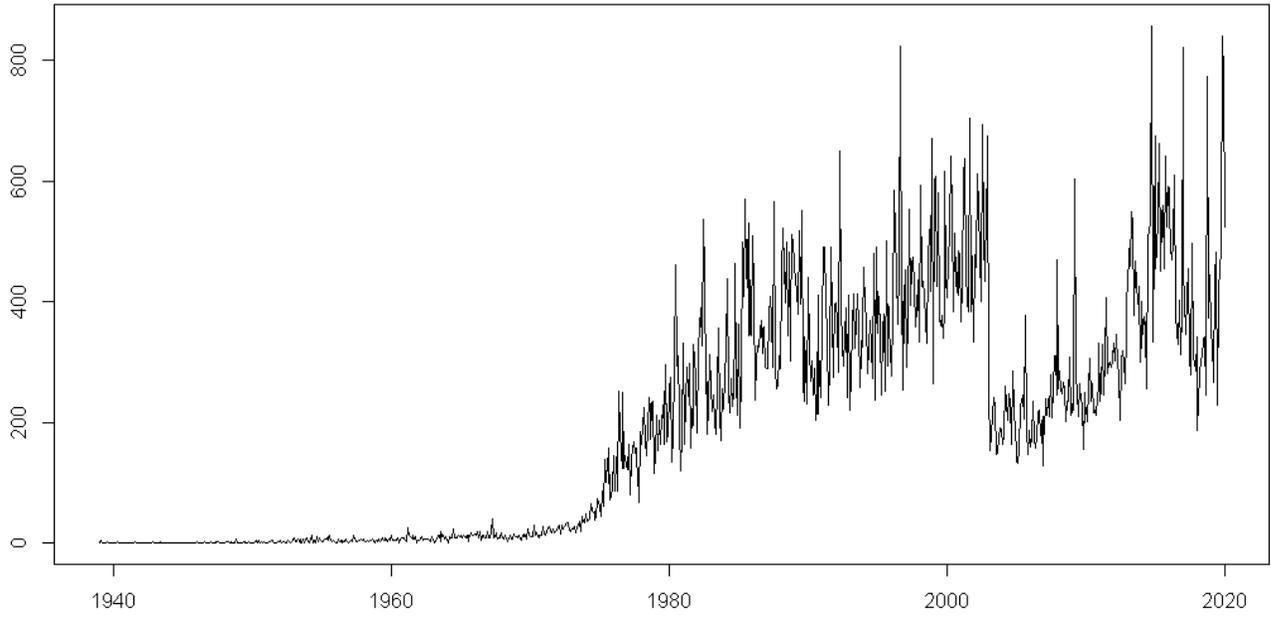

**Figure 2: Serial and partial correlation functions: lag is presented in months**

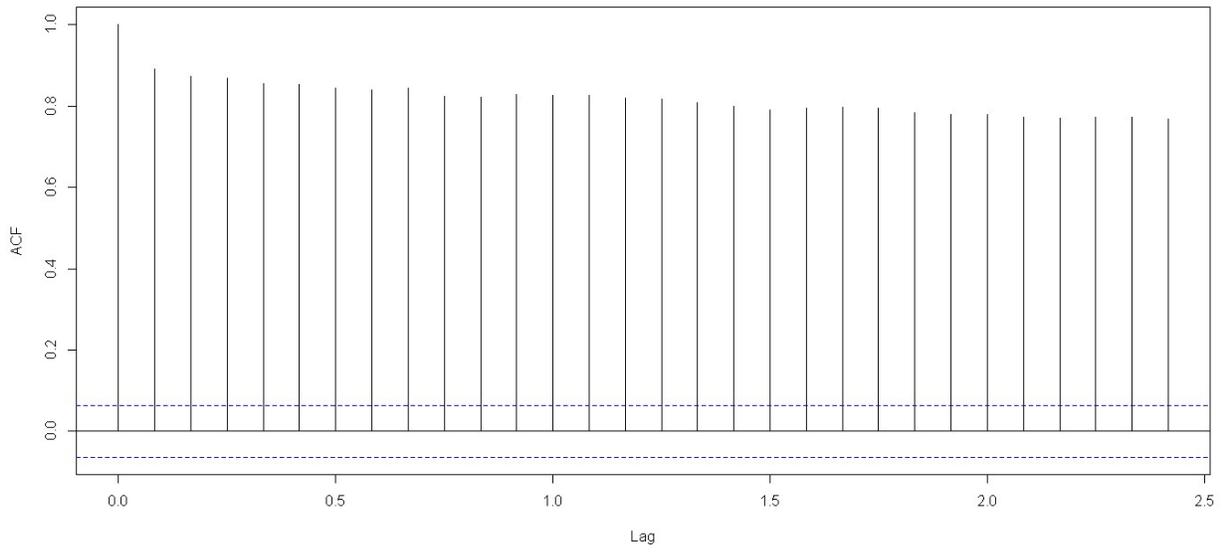

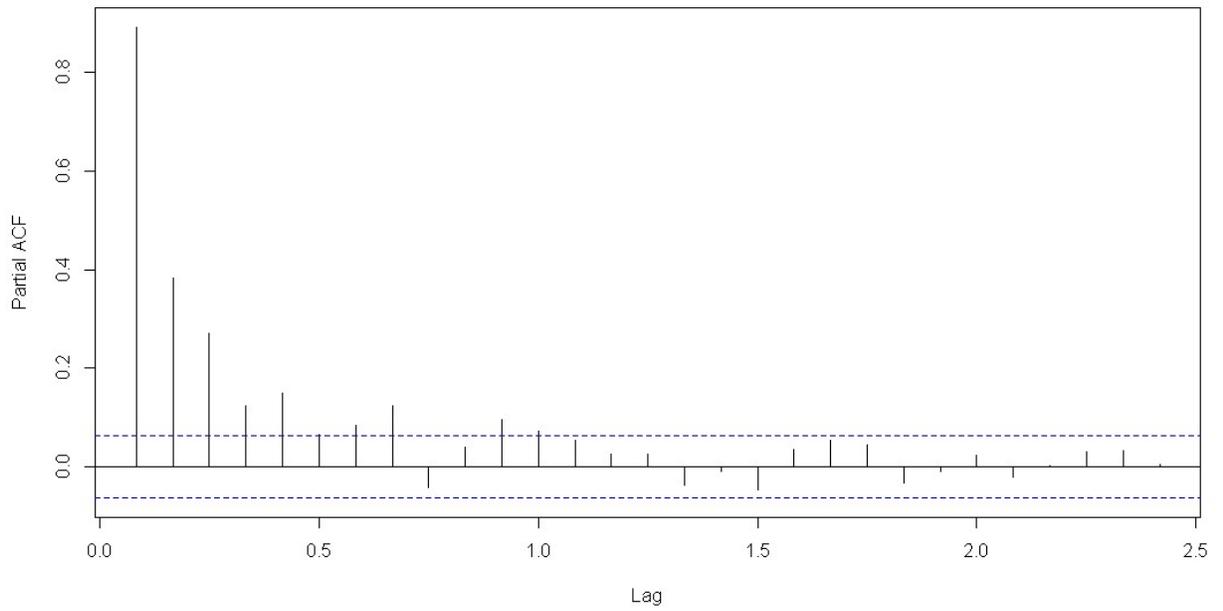

**Figure 3: The Chronological Hurst Exponent based on US Drug Approvals (Figure 1) from 1939 to 2019**

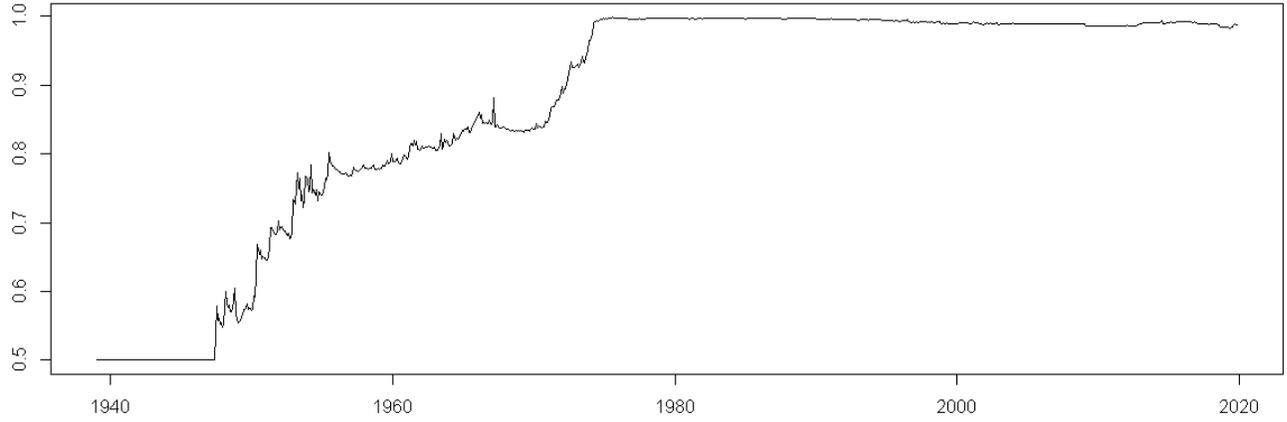

**Figure 4: Wavelet periodogram of US approvals: black lines are the wavelet power ridges and white contour lines to border the area of wavelet power significance of 99%**

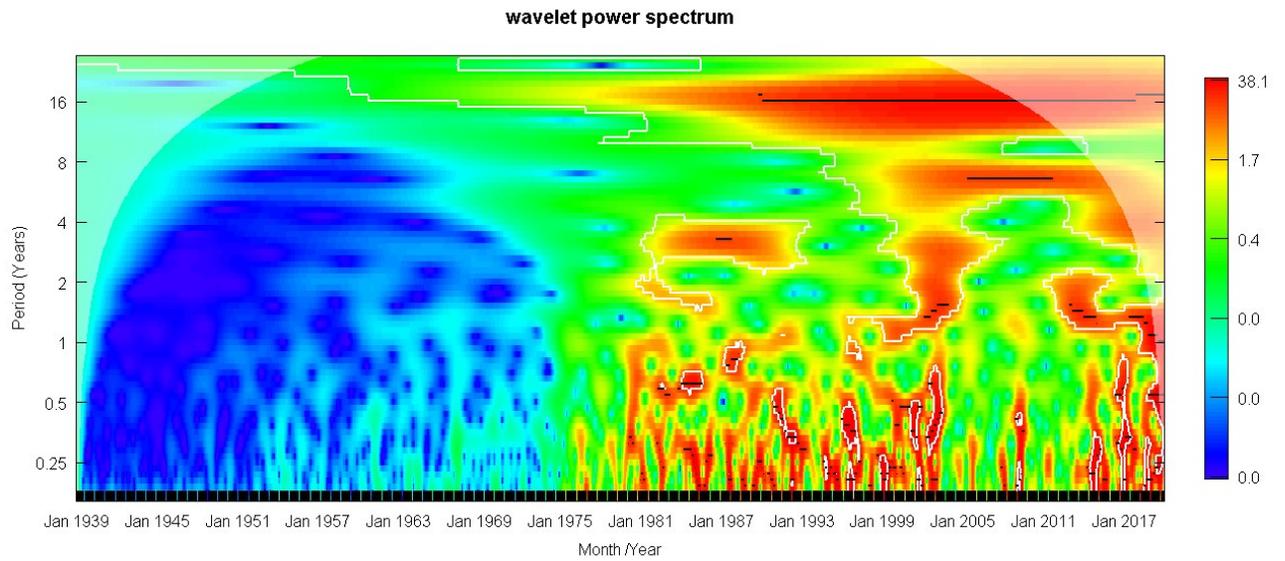

**Figure 5: Wavelet period versus power with 95% significant levels in red**

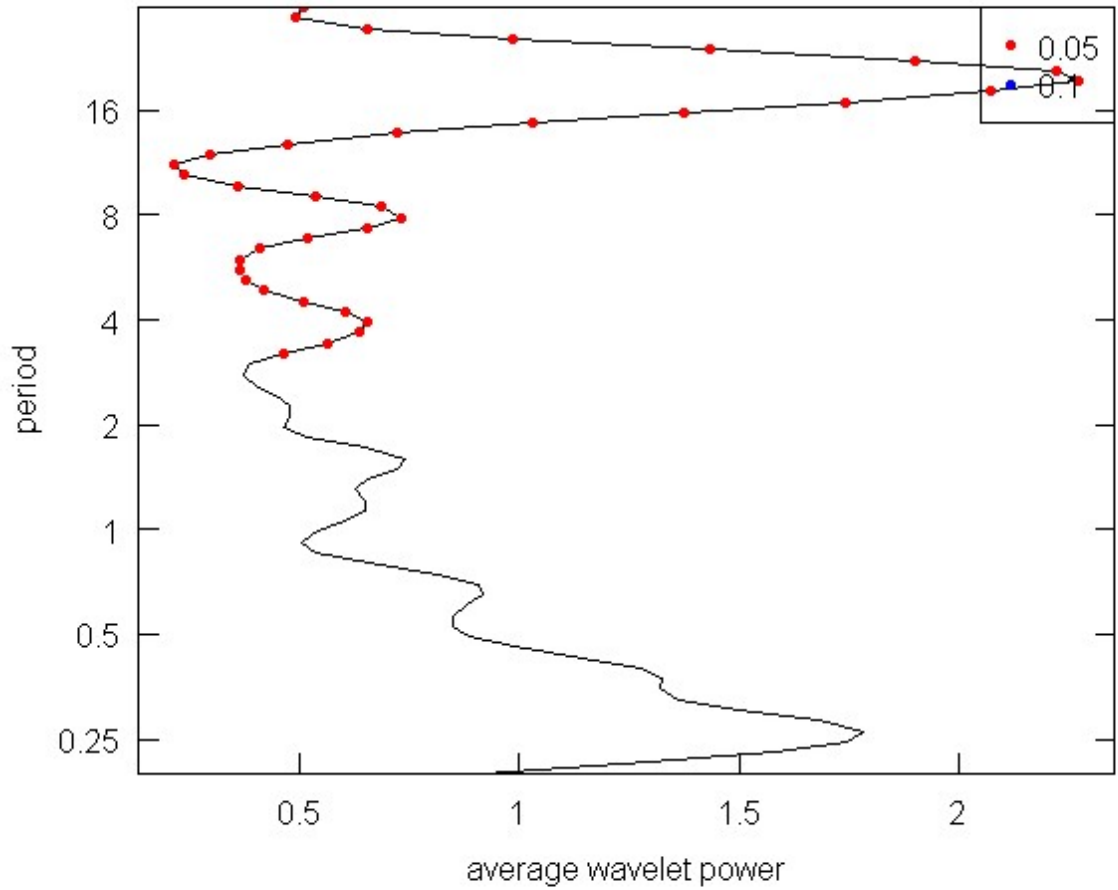

**Supplementary Materials**

### I. Data Collection

The FDA website https://www.accessdata.fda.gov/scripts/cder/daf/ was access on July 16 and July 17, 2020. The data was culled from a monthly report and described as follows (see Figure 1):

> "All Approvals and Tentative Approvals by Month.
>
> Reports include only BLAs/NDAs/ANDAs or supplements to those applications approved by the Center for Drug Evaluation and Research (CDER) and tentative NDA/ANDA approvals in CDER. The reports do not include applications or supplements approved by the Center for Biologics Evaluation and Research (CBER).
>
> Approvals of New Drug Applications (NDAs), Biologics License Applications (BLAs), and Abbreviated New Drug Applications (ANDAs), and supplements to those applications; and tentative approvals of ANDAs and NDAs."

Upon entry into the data-repository via the website, the number of approvals from Jan. 1939 to Dec. 2019 was then determined by month (see Figure 2). The values were placed in Excel and then exported as a comma delimited CSV file for input into the data analysis routine.

Figure 1: The FDA web data-repository allowing search of drug approval reports as a function of month.

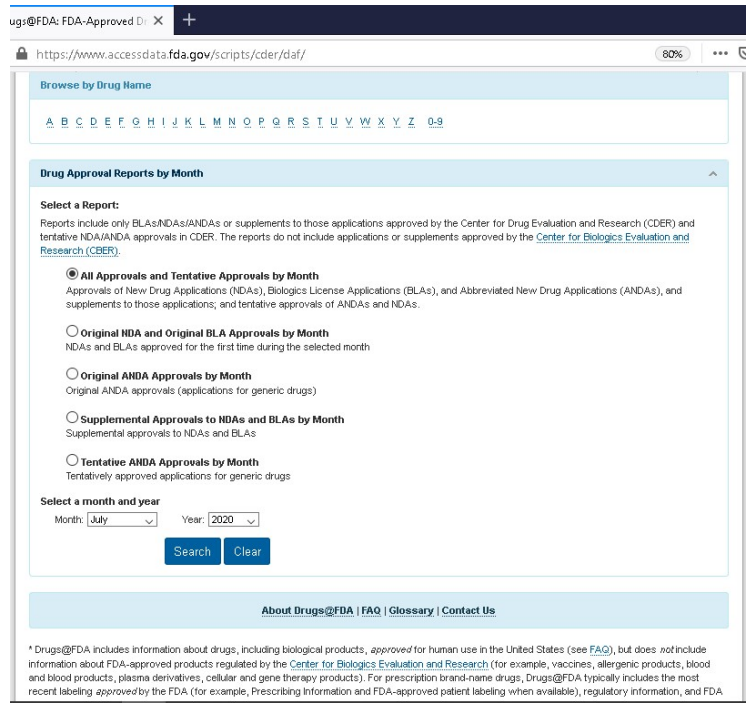

[Screenshot of Drugs@FDA: FDA-Approved Drugs webpage showing "All Approvals and Tentative Approvals February 1939" with four entries dated 02/09/1939 for HEPARIN SODIUM, LIQUAEMIN LOCK FLUSH, LIQUAEMIN SODIUM, and LIQUAEMIN SODIUM PRESERVATIVE FREE, all from ORGANON USA INC.]

**II.       Statistical Analysis**

Install R from: https://cloud.r-project.org/
citation()

> R Core Team (2020). R: A language and environment for
> statistical computing. R Foundation for Statistical
> Computing, Vienna, Austria. URL https://www.R-project.org/.

version

| | |
|---|---|
| platform | x86_64-w64-mingw32 |
| arch | x86_64 |
| os | mingw32 |
| system | x86_64, mingw32 |
| status | |
| major | 4 |
| minor | 0.2 |
| year | 2020 |
| month | 06 |
| day | 22 |
| svn rev | 78730 |
| language | R |
| version.string | R version 4.0.2 (2020-06-22) |
| nickname | Taking Off Again |

**#Step 1: Load data, convert to time series, perform descriptive statistics, and autocorrelation**

Input <- read.csv(file="c:\\Users/pzn6811/OneDrive - Takeda/Desktop/GLOC/read.csv", header=T, sep=",")

Input<-na.omit(Input) #excel seems to have some NAs at the end of column

time<-ts(Input$Number.of.Approvals,start=c(1939,1),end=c(2019,12),frequency=12)

time

```
     Jan Feb Mar Apr May Jun Jul Aug Sep Oct Nov Dec
1939   0   4   1   1   0   1   0   2   0   1   0   0
1940   0   0   0   2   0   0   0   0   1   0   0   0
1941   1   0   0   0   0   1   1   3   0   0   0   0
1942   0   1   0   0   1   0   0   0   0   0   2   1
1943   0   0   1   0   3   1   0   0   0   0   0   1
1944   0   0   0   1   1   0   1   0   0   0   0   1
1945   0   1   0   0   0   0   0   1   1   0   0
1946   3   1   1   1   1   0   0   2   0   1   1   0
1947   2   1   1   1   0   2   3   2   0   1   0   1
1948   0   3   2   2   0   0   2   0   1   1   6   0
1949   0   0   1   1   2   1   2   1   1   2   0   1
1950   1   0   2   4   1   5   4   1   2   0   2   1
1951   1   0   2   2   4   4   2   1   1   0   2   2
1952   5   1   2   2   0   2   0   4   2   0   2   4
1953   8   3   2   7   5   1   7   0   7   1   6   9
1954   4   0   5  12   1   5   3   1  11   2   9   4
1955   3   3   6   6   8   4   8  13   5   2   4   3
1956   0   1   6   2   4   1   4   3   2   7   3   1
1957   2   5   2  13   8   4   3   3   4   5   6   5
1958   6   2   2   4   2   5   4   4   6   5   0   4
1959   3   6   3   5   8   4   4   8   7   3   6   5
1960  14   4   5   5   5   8   2   1   5   8   8   8
1961   4   2  25  13  11   8   6  13   6  10   1   5
1962   4   7  11   5   7   7   6   6   7   4   3  10
```

1963   7   1   3   6  12   9  20   3  15  14   7  11
1964   7   1  10   8   9  24   9  11   9   9  14  11
1965  14   9  12  10   9  13   3  12  11  17  12  13
1966  15  13  19   8  20   5  13   9  12   9   8  20
1967   9   8  19  41   8  11  17  12   7   8   9  17
1968   9   6  12   9   3   8  10  13   6   7  16  10
1969  10   8  12   7   4  12  16  10  12   7  24  15
1970  10  11  11  30  13  17  11  13   9  13  14  27
1971  16  14  20  23  28  21  17  24  24  18  19  21
1972  26  30  15  30  23  28  29  33  34  34  21  20
1973  23  24  27  17  28  30  35  20  19  44  36  33
1974  48  37  38  40  66  55  50  52  39  43  73  70
1975  62  44  87  61 138 102 140  98 157  72  77  90
1976 145  87 143  87 251 185 149 124 250 124 145 128
1977 121 164  81 139 158 169 150 158 131 120  68 185
1978 164 193 224 170 144 190 242 223 172 230 234 116
1979 145 213 153 203 164 213 208 252 163 295 168 180
1980 254 275 135 179 290 462 293 310 219 191 119 178
1981 331 163 238 292 243 297 158 222 195 329 309 273
1982 183 323 356 391 328 536 449 247 267 180 224 312
1983 261 218 246 201 180 263 356 210 170 176 256 223
1984 274 322 439 247 217 272 226 249 359 463 270 211
1985 362 190 276 498 408 570 438 503 344 530 347 344
1986 509 421 238 303 328 326 353 369 314 354 359 292
1987 289 290 378 408 375 291 565 287 256 271 260 310
1988 290 446 522 459 399 498 344 482 303 360 511 498
1989 463 434 422 379 518 397 551 262 236 341 301 231
1990 441 323 269 303 245 290 203 222 410 214 302 241
1991 363 491 490 404 339 229 294 490 375 274 300 395
1992 399 367 283 649 395 326 294 343 389 241 283 412

Persistency and cyclicity in US drug approvals                                    Author: Daizadeh, I.1993 221 278 361 413 324 365 413 360 258 335 290 359

1994 458 394 371 282 323 370 325 273 480 327 237 491

1995 349 409 355 246 356 379 252 501 294 392 356 321

1996 277 377 584 512 451 362 456 824 346 401 254 405

1997 452 292 366 554 395 462 473 425 359 391 386 436

1998 334 593 425 435 411 369 332 468 520 476 418 670

1999 265 601 607 426 581 373 361 377 339 385 617 439

2000 406 468 516 642 550 383 513 437 416 485 410 480

2001 366 428 607 638 452 398 383 704 383 475 459 333

2002 479 457 613 555 477 400 693 506 434 569 607 674

2003 255 153 179 214 242 227 147 150 176 192 177 189

2004 164 177 261 232 204 247 232 163 286 226 179 201

2005 136 133 154 227 235 245 205 378 187 146 183 186

2006 160 173 235 172 158 176 216 221 178 204 129 209

2007 181 240 212 240 209 279 207 256 310 311 245 469

2008 257 286 257 248 266 257 202 234 236 308 211 250

2009 217 221 604 278 250 210 264 235 233 156 250 199

2010 224 202 255 307 234 264 216 221 211 243 259 331

2011 235 255 330 246 306 407 282 297 301 292 300 309

2012 331 303 346 301 280 204 302 319 302 265 350 413

2013 417 488 460 549 533 377 468 431 396 425 299 343

2014 400 346 376 361 255 509 514 537 858 334 507 675

2015 456 490 535 662 451 545 560 454 641 524 592 589

2016 556 488 470 496 610 378 350 329 402 313 384 821

2017 449 405 345 411 455 306 280 497 307 299 285 342

2018 186 233 281 303 310 337 341 246 773 374 444 356

2019 342 338 266 441 483 229 419 453 488 840 776 524

plot(time)

Page 35 of 45

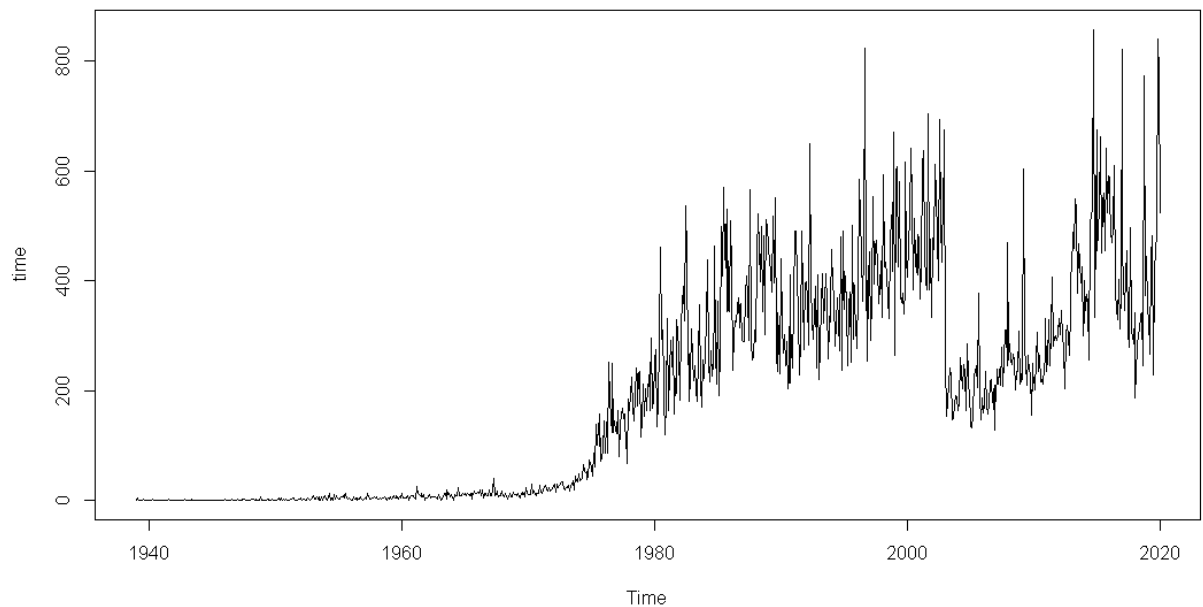

summary(time)

> Min. 1st Qu. Median Mean 3rd Qu. Max.
> 0.0  5.0  164.0  186.4  329.2  858.0

library(moments)
citation("moments")

> Lukasz Komsta and Frederick Novomestky (2015). moments: Moments, cumulants, skewness, kurtosis and related tests. R package version 0.14. https://CRAN.R-project.org/package=moments

sd(time)

> 190.9333

Kurtosis(time) #platykurtic (excess kurtosis = kurtosis – 3)

> 2.598539

Skewness(time)

> 0.6980762

acf(time);pacf(time)

ndiffs(time)

> [1]    1

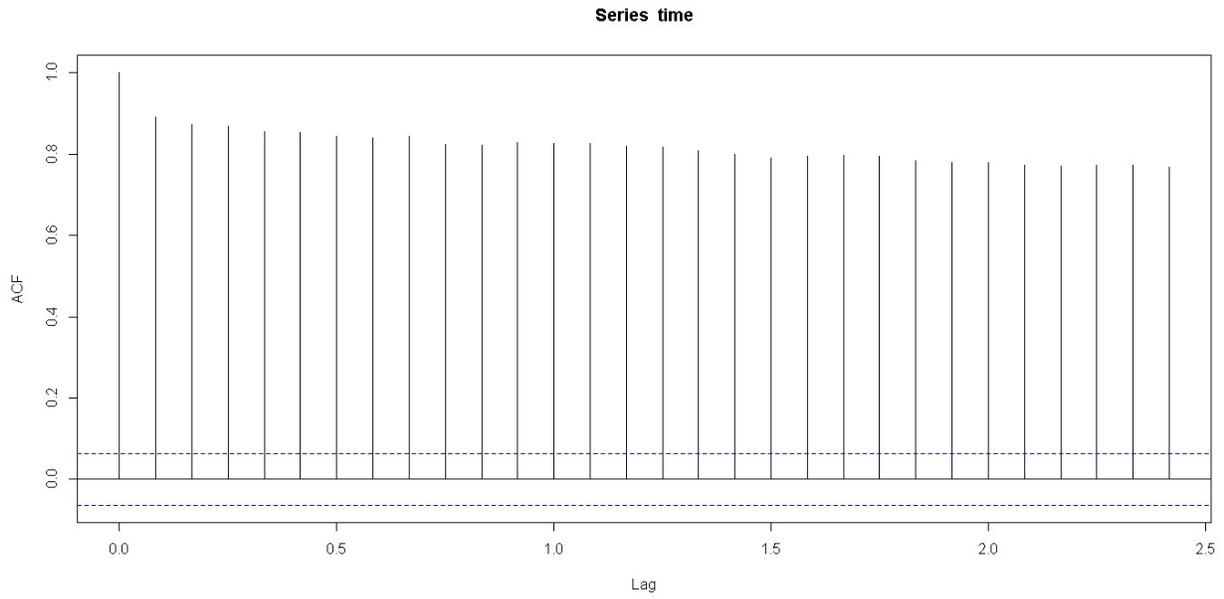

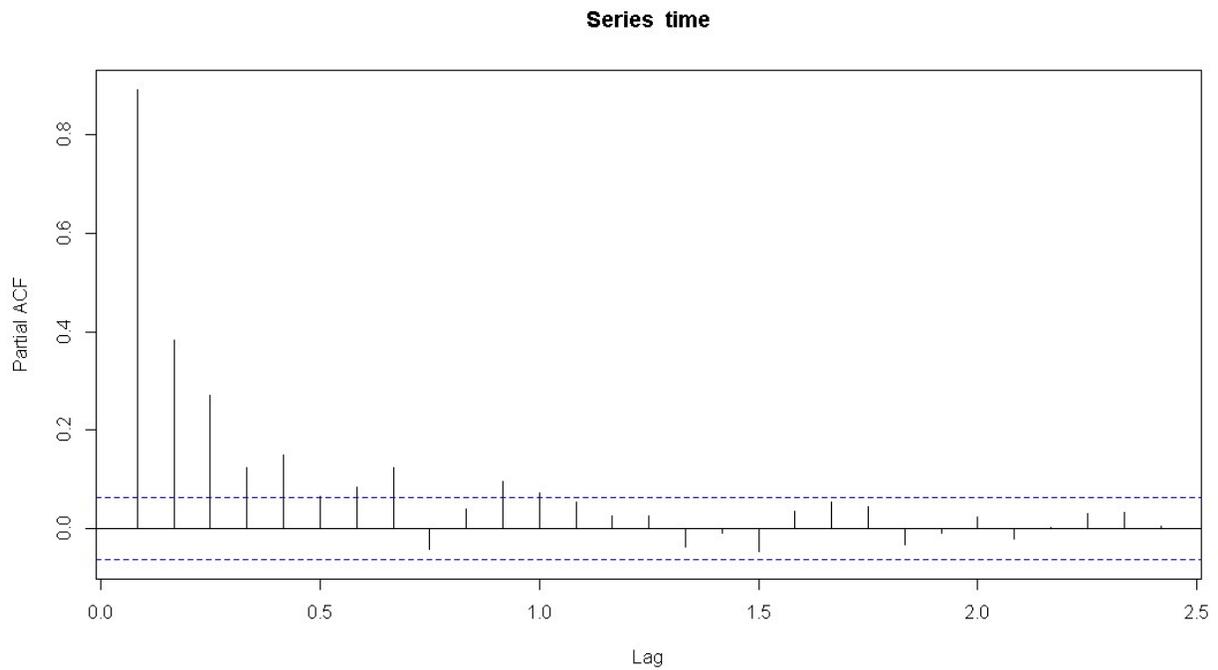

#Step 2: Perform normality, stationarity, seasonality, long-memory, and non-linearity tests

#normality test

library(nortest) #all normality tests rejected hypothesis of normality – presenting two

citation("nortest")

> Juergen Gross and Uwe Ligges (2015). nortest: Tests for Normality. R package version 1.0-4.
> https://CRAN.R-project.org/package=nortest

ad.test(time) #null normality

> Anderson-Darling normality test
> data:  time
> A = 48.166, p-value < 2.2e-16

Cvm.test(time)

> Cramer-von Mises normality test
> data:  time
> W = 7.5428, p-value = 7.37e-10
> Warning message:
> In cvm.test(time) :
> p-value is smaller than 7.37e-10, cannot be computed more accurately

#stationarity test

Library(aTSA)

Citation("aTSA")

> Debin Qiu (2015). aTSA: Alternative Time Series Analysis. R package
> version 3.1.2. https://CRAN.R-project.org/package=aTSA

stationary.test(time,method="kpss")

> KPSS Unit Root Test
> alternative: nonstationary
> Type 1: no drift no trend
> lag stat p.value
> 7 6.32   0.01
> -----
> Type 2: with drift no trend
> lag stat p.value
> 7   7   0.01
> -----
> Type 1: with drift and trend
> lag  stat p.value
> 7 0.671   0.01
> -----------
> Note: p.value = 0.01 means p.value <= 0.01
> : p.value = 0.10 means p.value >= 0.10

stationary.test(diff(time),method="kpss")

> KPSS Unit Root Test
> alternative: nonstationary

       Type 1: no drift no trend
       lag   stat p.value
       7 0.0776    0.1
       -----
       Type 2: with drift no trend
       lag   stat p.value
       7 0.0281    0.1
       -----
       Type 1: with drift and trend
       lag   stat p.value
       7 0.0162    0.1
       -----------
       Note: p.value = 0.01 means p.value <= 0.01
       : p.value = 0.10 means p.value >= 0.10

#long-memory test

library(LongMemoryTS)

citation("LongMemoryTS")

       Christian Leschinski (2019). LongMemoryTS: Long Memory Time Series. R package version 0.1.0.
       https://CRAN.R-project.org/package=LongMemoryTS

m<-floor(1+500^0.75)

# Qu test

       Qu.test(diff(Input$Number.of.Approvals),m)
       $W.stat
       [1] 1.033545

       $CriticalValues

              eps=.02 eps=.05
       alpha=.1     1.118   1.022
       alpha=.05    1.252   1.155
       alpha=.025   1.374   1.277
       alpha=.01    1.517   1.426

#Multivariate local Whittle Score

MLWS(diff(Input$Number.of.Approvals), m=m)

     $B
     [,1]
     [1,] 1

     $d
     [1] 0.9172231

```
$W.stat
[1] 0.9172231

$CriticalValues
alpha=.1  alpha=.05 alpha=.025  alpha=.01
   1.11`8     1.252     1.374     1.517
```

#Seasonality tests

library(seastests)

citation("seastests")

> Daniel Ollech (2019). seastests: Seasonality Tests. R
> package version 0.14.2. https://CRAN.R-project.org/package=seastests

#Webel-Ollech overall seasonality test
summary(wo(time))

> Test used:  WO
> Test statistic:  0
> P-value:  1 1 0.05105411
> The WO - test does not identify seasonality

#calculate through variety of tests
isSeasonal(time, "qs")  #QS test

[1] FALSE

isSeasonal(time, "fried") #Friedman test

[1] FALSE

isSeasonal (time, "welch") #Welch test

[1] FALSE

#Nonlinearity tests

library(nonlinearTseries)

citation("nonlinearTseries")

> Constantino A. Garcia (2020). nonlinearTseries: Nonlinear Time Series Analysis. R package
> version 0.2.10. https://CRAN.R-project.org/package=nonlinearTseries

> nonlinearityTest(time)

>  ** Teraesvirta's neural network test  **
> Null hypothesis: Linearity in "mean"
>  X-squared =  227.9227  df =  2  p-value = 0

```
       ** White neural network test  **
Null hypothesis: Linearity in "mean"
X-squared =  227.1936  df =  2  p-value =  0

       ** Keenan's one-degree test for nonlinearity  **
Null hypothesis: The time series follows some AR process
F-stat =  17.08669  p-value =  3.888728e-05

       ** McLeod-Li test  **
Null hypothesis: The time series follows some ARIMA process
Maximum p-value =  0

       ** Tsay's Test for nonlinearity **
Null hypothesis: The time series follows some AR process
F-stat =  2.733688  p-value =  6.342547e-14

      ** Likelihood ratio test for threshold nonlinearity **
      Null hypothesis: The time series follows some AR process

     Alternative hypothesis: The time series follows some TAR process
     X-squared =  47.58834  p-value =  0.0004552571
```

**#Step 3: Develop Hurst over time**

library(tsfeatures)
citation("tsfeatures")

> Rob Hyndman, Yanfei Kang, Pablo Montero-Manso, Thiyanga Talagala, Earo Wang, Yangzhuoran Yang and Mitchell O'Hara-Wild (2020). tsfeatures: Time Series Feature Extraction. R package version 1.0.2. https://CRAN.R-project.org/package=tsfeatures

hurstApprovals=0

end<-length(time)

for (i in 1:end) { hurstApprovals[i]  <- hurst (time[1:(1+i*1)]) }

hurstApprovals<-ts(hurstApprovals,start=c(1939,1),end=c(2019,12),frequency=12)

plot(hurstApprovals)

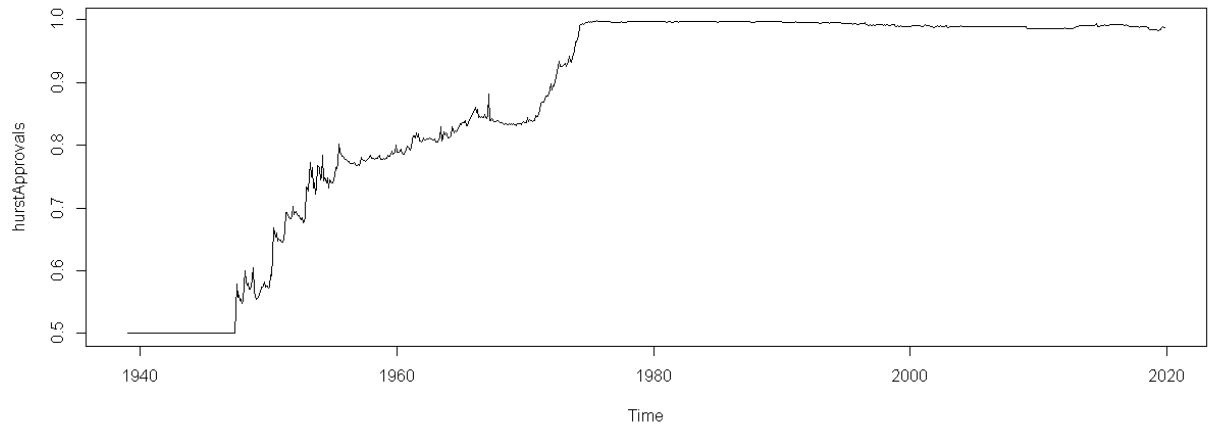

**#Identify periods**

**#Method 1: The Wavelet Power Spectrum Of A Single Time Series**
#Note: Loess smoothing as default is 0.75 for this parameter

library(WaveletComp)

citation("WaveletComp")

> Angi Roesch and Harald Schmidbauer (2018). WaveletComp: Computational Wavelet Analysis. R package version 1.1. https://CRAN.R-project.org/package=WaveletComp

monthyear <- seq(as.Date("1939-01-01"), as.Date("2019-12-31"), by = "month")
monthyear <- strftime(monthyear, format = "%b %Y")
c<- analyze.wavelet(data.frame(time),"time", dt=1/12, dj=0.1)
wt.image(c,  main = "wavelet power spectrum", periodlab = "Period (Years)", timelab = "Month /Year",
spec.time.axis = list(at = 1:length(monthyear), labels = monthyear))

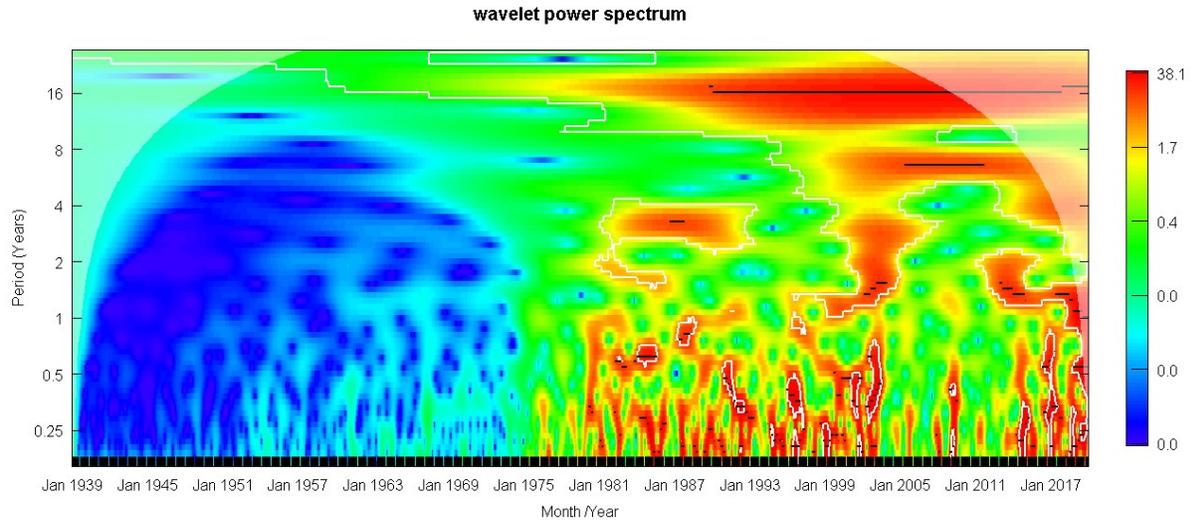

wt.avg(c)

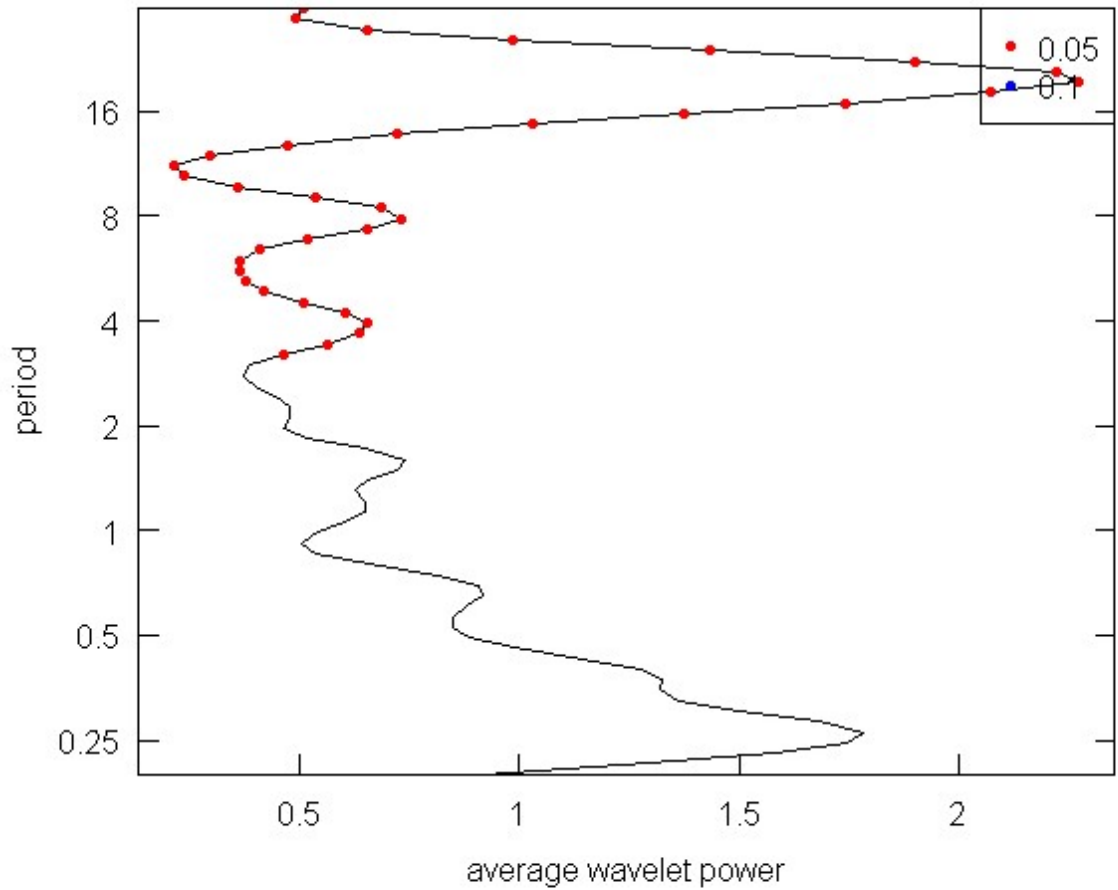

#Method 2: Continuous Morlet Wavelet Transform

Library(dplR);citation("dplR")

> Bunn AG (2008). "A dendrochronology program library in R (dplR)."_Dendrochronologia_, *26*(2), 115-124. ISSN 1125-7865, doi:10.1016/j.dendro.2008.01.002 (URL: https://doi.org/10.1016/j.dendro.2008.01.002).

> Bunn AG (2010). "Statistical and visual crossdating in R using the dplR library." _Dendrochronologia_, *28*(4), 251-258. ISSN 1125-7865, doi: 10.1016/j.dendro.2009.12.001 (https://doi.org/10.1016/j.dendro.2009.12.001).

> Andy Bunn, Mikko Korpela, Franco Biondi, Filipe Campelo, Pierre Mérian, Fares Qeadan and Christian Zang (2020). dplR: Dendrochronology Program Library in R. R package version 1.7.1. https://CRAN.R-project.org/package=dplR

wave.out <- morlet(time, p2 = 8, dj = 0.1, siglvl = 0.95)

wave.out$period <- wave.out$period/12

wavelet.plot(wave.out)

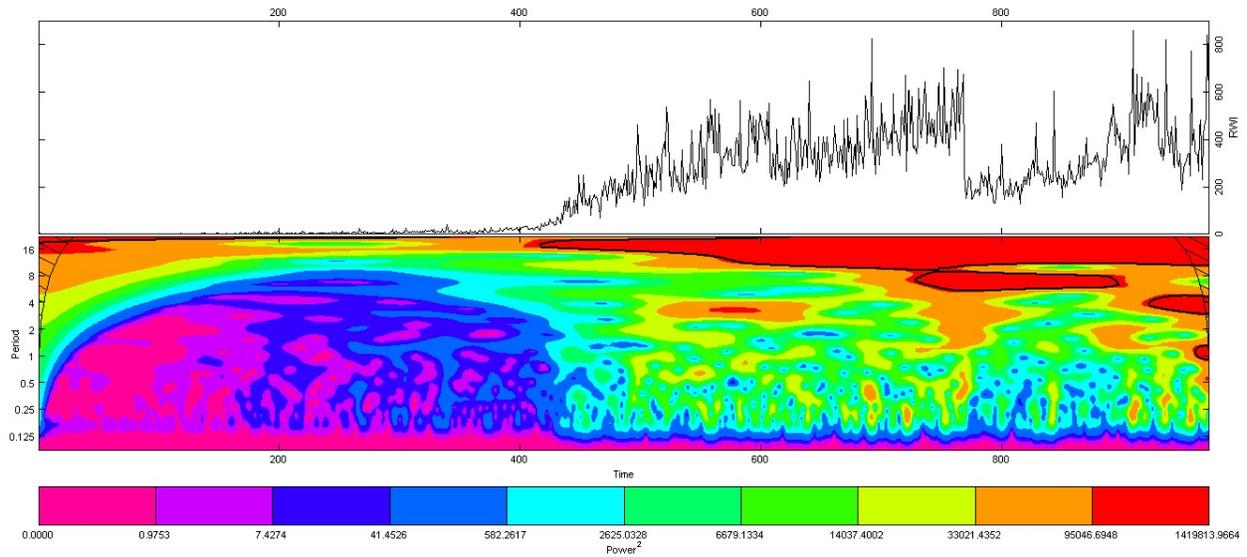

wave.avg <- data.frame(power = apply(wave.out$Power, 2, mean), period = (wave.out$period))

plot(wave.avg$period, wave.avg$power, type = "l")

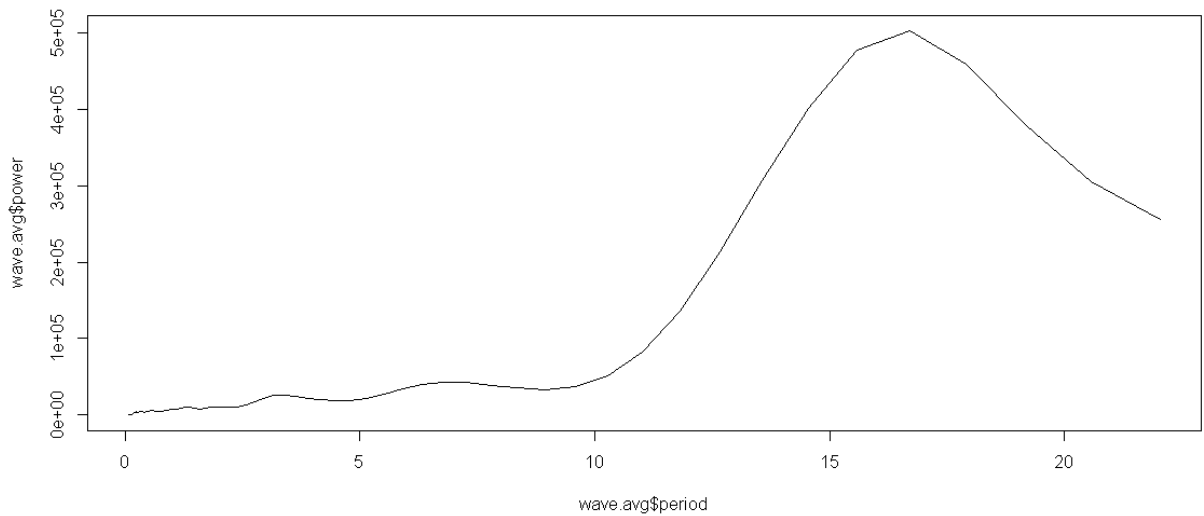

#Confirm time series frequency

library(forecast);citation("forecast")

findfrequency(time) # dominant frequency is determined from a spectral analysis of the time series

[1]     17